\documentstyle[preprint,revtex]{aps}

\begin{document}
\draft
\preprint{\flushright KEK-TH-345\\
              KEK Preprint 92-87\\
              November 1992\\}
\begin{title}
 Chiral Structure of the $b\rightarrow c$ Charged Current
                    \\and\\
         the Semileptonic $\Lambda_b$ Decay
\end{title}
\author{Minoru Tanaka\cite{JSPS}\cite{EMA}}

\begin{instit}
Theory Group, National Laboratory for High Energy Physics (KEK),\\
Oho 1-1, Tsukuba, Ibaraki 305, Japan
\end{instit}

\begin{abstract}
Using the heavy quark effective theory, chiral structure of
the $b\rightarrow c$ charged current can be determined by
the semileptonic $\Lambda_b$ decay with satisfactory
theoretical accuracy. We define an asymmetry which is sensitive to
the chirality of the $b\rightarrow c$ charged current. We show
that this asymmetry has no theoretical uncertainty in
the heavy quark limit. It is also shown
that $1/m_c$ correction to this asymmetry is small and disappears
at the kinematical point of zero momentum transfer.
\end{abstract}

\narrowtext

\section{Introduction}
The most conspicuous feature of the weak charged current in
the standard model is its pure left-handedness. It is
implemented by the transformation properties
of the chiral fermions: The left-handed fermions are assigned
to doublets of the weak $SU(2)$ and the right-handed ones
are to singlets.

Another important aspect of the charged current is the mixing.
In the standard model, there are three generations
of quarks and leptons, and these quarks mix with each other,
while the mixing of the leptons has no
physical meanings. This quark mixing leads to a $3 \times 3$ matrix
(Kobayashi-Maskawa matrix \cite{KM}) in the quark charged current,
resulting in nine components of the charged current in
the quark sector. The quark mixing does not alter
the above-mentioned chiral structure of the charged
current in the standard model. In other words,
the standard model predicts
that all nine quark flavor (and three leptonic)
components of the weak charged
current have the identical chiral structure, pure left-handedness.

This prediction, however, may be altered if extended
models are considered.
For example, in the Left-Right gauge models based on the
$SU(2)_L\times SU(2)_R\times U(1)$ weak gauge group \cite{LR},
the exchange of the right-handed $W$ boson and
the mixing between the left-handed
$W$ boson and the right-handed one cause the
right-handed component in the charged current. If the model is
manifestly Left-Right symmetric, the above flavor universality
of the chiral structure of the charged current is maintained
although the pure left-handedness is lost.

While, even the flavor universality is not preserved
in the non-manifest models because of the difference
between the left-handed quark mixing matrix and
its right-handed counterpart.
In this kind of models, the mass of the right-handed $W$
boson can be relatively light ($\sim$a few hundred GeV) and
non-negligible part of some flavor components of the charged
current may be caused by the right-handed interaction~\cite{OE}.

Recently, an extreme model along this line was proposed
by Gronau and Wakaizumi \cite{GWPRL}.
In their model, $b$ quark decay is caused purely
by a right-handed current.

One can suppose another example in which both the pure
left-handedness and the flavor universality are broken.
It is the standard model with the vector-like fourth generation.
This model also contains the right-handed charged current
which originates from the $SU(2)$ gauge interaction of
the right-handed doublet fermions in the fourth generation.
An interesting speculation on this kind of model was made by
Fritzsch \cite{F}. Following his speculation,
the phenomenon of the parity violation is interpreted as
a low energy phenomenon and related to the smallness of
the usual quark and lepton masses. The parity violation becomes
maximal in the limit of vanishing fermion masses.
Therefore, the magnitude of the right-handed charged current
is expected to be larger in heavier quark sectors.
This means the loss of the flavor universality of the chiral
structure of the charged current. Actually, it is shown in
ref.~\cite{F} that the strength of the right-handed charged
current which consists of $q$ and $q'$ quarks is proportional
to the factor $\sqrt{m_q m_{q'}}$.

Following the above arguments, the chiral structure of all the nine
quark (and three leptonic) charged currents should be
examined separately. It, however, is not straightforward
to determine the chirality of the {\em quark} charged
currents experimentally, because we observe
{\em hadrons} not quarks. We have to extract the quark
couplings from the observed hadronic
interaction. In spite of this difficulty, the chiral structures of
the charged currents which contain only the light quarks are
well constrained to be left-handed. Actually, the result of
inelastic neutrino interaction \cite{CDHS1} suggests
$\vert g_R/g_L \vert\alt 0.1$ for the $d\rightarrow u$
charged current, and the PCAC consideration on
the $K\rightarrow 2\pi$ and
$K\rightarrow 3\pi$ decays gives limits as
$\vert g_R/g_L \vert\alt 0.004$ for
the $d\rightarrow u$ and $s\rightarrow u$
charged current \cite{DH}.

On the other hand, since we cannot use the chiral symmetry
for heavy quarks ($c$, $b$ and $t$) in addition to
the difficulty of experiments like deep inelastic scattering
due to the CKM suppression,
there are poorer evidences of the chiralities of the charged
currents involving the heavy quarks. In fact,
the experimental bound on the $c\rightarrow d,s$ currents
is $\vert g_R/g_L \vert\alt 0.3$ \cite{CDHS2}, and
$\vert g_R/g_L \vert\alt 0.5$ for $b\rightarrow c$ current
has been given recently by CLEO using
$B\rightarrow D^*\ell\nu$ decay \cite{CLEO}.
We have no direct limits on the other charged
currents involving $b$ and/or $t$. It should be noticed that
the test made by CLEO assumed the left-handedness of the leptonic
charged current, and it cannot discriminate the model proposed in
ref.~\cite{GWPRL} from the standard model.
In this sense, this test is incomplete \cite{GWPRL,GWPL}.

{}From the theoretical point of view, recently a remarkable
progress has been made in treating hadrons which
contain a quark of much heavier mass than the QCD scale.
It was pointed out that an effective theory of QCD with
$N_h$ heavy quarks has new symmetry $SU(2N_h)$, associated with
the flavor and the spin rotations of the heavy quarks \cite{IW}.
Such a theory is called as the heavy quark effective theory.
This symmetry, especially its spin rotation part, is expected
to play a crucial role to determine the chiral structure
of the charged current involving the heavy quarks.

In this paper, we discuss a method to detect the effect
of the right-handed $b\rightarrow c$ current in the
semileptonic decay of $\Lambda_b$,
which is supposed to be the lightest bottom baryon,
into $\Lambda_c$.
In the following analysis, we utilize the heavy quark effective
theory where we regard both $b$ and $c$ quarks as heavy.
The bottom decay is caused by this current for the most part
and its detailed study may lead us to something beyond
the standard model.

Moreover, we stress that the $\Lambda_{b,c}$ baryons are
simpler systems than the $B^{(*)}$ and $D^{(*)}$ mesons in
a standpoint of the heavy quark effective theory,
because the light degrees of freedom form a zero-spin system.
The simplicity of these baryons leads to the most
important result of our analysis that the theoretical
uncertainty is small enough to see the chirality of
the $b\rightarrow c$ current. The $1/m_c$
correction is known to be controlled by one dimensionful
parameter \cite{GGW}, and we found that, as for the asymmetry
which we will define in the following section,
even the effect of $1/m_c$ correction through
this parameter vanishes at the kinematical point of
{\em zero} momentum transfer.

In section 2, we present our formalism of the differential
decay rate, and define an asymmetry which is sensitive to
the chirality of the $b\rightarrow c$ current.
Section 3 includes the implication of the heavy quark limit
and the numerical result in this limit. In section 4,
the $1/m_c$ correction is discussed. We state two remarks
and our conclusion in section 5.

\section{Formalism}

In this section, we present an expression of a double
differential decay rate of
$\Lambda_b\rightarrow \Lambda_c \ell \bar\nu$ decay and
define an asymmetry which is a good probe of the chirality
of the $b\rightarrow c$ charged current.
Here we assume a polarization of the initial $\Lambda_b$
and sum up with the final state spins.
The initial polarization is realized in
$e^+e^-\rightarrow Z \rightarrow b \bar b$
followed by the hadronization $b\rightarrow \Lambda_b$ \cite{MS}.
Note that the semileptonic $\Lambda_b$ decay in
this process has already been observed at LEP \cite{LEP}.
We work with this production process of $\Lambda_b$ in mind,
and therefore we choose the frame in which the initial
$\Lambda_b$ is running \cite{RWGW}.

We start with the definition of form factors which can appear in
$\Lambda_b\rightarrow\Lambda_c$ transition by weak vector and
axial-vector currents:

\begin{equation}
\left\langle \Lambda_c(v',s')\right\vert
\bar c \gamma_\mu b \left\vert \Lambda_b(v,s)\right\rangle
\equiv
\bar u_{\Lambda_c}(v',s')(F_1\gamma_\mu+F_2v_\mu+F_3v'_\mu)
     u_{\Lambda_b}(v,s)\;,
\label{VEC}
\end{equation}

\begin{equation}
\left\langle \Lambda_c(v',s')\right\vert\bar c \gamma_\mu \gamma_5 b
\left\vert \Lambda_b(v,s)\right\rangle\equiv
\bar u_{\Lambda_c}(v',s')(G_1\gamma_\mu+G_2v_\mu+G_3v'_\mu)\gamma_5
u_{\Lambda_b}(v,s)\;,
\label{AX}
\end{equation}
where $v=p_b/m_{\Lambda_b}$ and $v'=p_c/m_{\Lambda_c}$ are
the four-velocity of the $\Lambda_b$ and $\Lambda_c$
respectively, and $F_i$'s and $G_i$'s are functions of $v\cdot v'$.

Here, we consider the differential decay rate of
$\Lambda_b\rightarrow \Lambda_c \ell \bar\nu$ with respect to
the energy of $\Lambda_c$ ($E_c$) and the momentum transfer
squared ($q^2=(p_b-p_c)^2$). One can write this double
differential rate as
\begin{equation}
\frac{d{\mit\Gamma}}{dx_c dq^2}=
J(q^2)+{\cal P}K(q^2)(x_c-\bar x_c)\;,
\label{DDR}
\end{equation}
where $x_c=E_c/E_b$, $\bar x_c=p_b\cdot p_c /m_{\Lambda_b}^2$,
$E_b$ is the energy of the initial $\Lambda_b$,
and $\cal P$ is the initial $\Lambda_b$ polarization which is,
as will be explained, equal to that of the initial $b$ quark
in the limit of infinite $m_b$.
The polarization of $b$ quark which comes from the above mentioned
process $Z\rightarrow b\bar b$ is given by
\begin{equation}
{\cal P}=\frac{2G_VG_A}{G_V^2+G_A^2}\simeq -0.93\,,
\label{P}
\end{equation}
where $G_V$ and $G_A$ are the vector and the axial-vector
coupling constants in $Zb\bar b$ vertex respectively,
and we used $\sin^2\theta_W=0.233$.
Note that the functions $J(q^2)$
and $K(q^2)$ are independent of $x_c$.
The range of $x_c$ is given by $x_c^{min}\leq x_c\leq x_c^{max}$ with
\begin{equation}
x_c^{max,min}=\bar x_c\pm \frac{\lambda}{2}\;,
\quad \lambda=\beta\frac{\sqrt{w_{+} w_{-}}}
{m_{\Lambda_b}^2}\;,
\end{equation}
where $\beta=\vert \mbox{\protect\boldmath $p$}_b \vert/E_b$
is the rapidity of $\Lambda_b$ and
$w_\pm=(m_{\Lambda_b}\pm m_{\Lambda_c})^2-q^2$.

$J(q^2)$ and $K(q^2)$ are written in terms of
the couplings of the
$b\rightarrow c$ vector and axial-vector current
(denoted by $g_V$ and $g_A$ respectively \cite{FOOT1}),
and the form factors $F_i$'s and $G_i$'s.
Their explicit forms are given in the appendix.
Note that the leptonic couplings
$f_V$ and $f_A$ appear in an overall factor,
since we neglected the lepton mass and integrated the lepton
system completely.

A genuine parity violating effect in the $b\rightarrow c$
charged current is expected to be observed through
the correlation between the spin of
$\Lambda_b$ and the momentum of $\Lambda_c$. Since the second term
in eq.~(\ref{DDR}) expresses this correlation, an energy asymmetry
which picks it up is defined as
\begin{eqnarray}
A(q^2) &\equiv & \frac{\displaystyle
  \int_{\bar x_c}^{x_c^{max}}\frac{d{\mit\Gamma}}{dx_c dq^2}dx_c-
  \int^{\bar x_c}_{x_c^{min}}\frac{d{\mit\Gamma}}{dx_c dq^2}dx_c}
                       {\displaystyle
  \int_{\bar x_c}^{x_c^{max}}\frac{d{\mit\Gamma}}{dx_c dq^2}dx_c+
  \int^{\bar x_c}_{x_c^{min}}\frac{d{\mit\Gamma}}{dx_c dq^2}dx_c}
                   \label{AD}\\
    & = & \frac{\lambda}{4}{\cal P}\frac{K(q^2)}{J(q^2)}\;.\label{A}
\end{eqnarray}
Note that we do not integrate with $q^2$ in the r.~h.~s. of
eq.~(\ref{AD}) because $J(q^2)$ and $K(q^2)$
involve the unknown form factors in eqs.~(\ref{VEC}) and (\ref{AX}).
As explained below, however, $A(q^2)$ itself involves
no unknown functions in the heavy quark limit, because all
form factors are proportional to an unknown
function in this limit \cite{IWMRR}. Moreover, it is worth
while pointing out that
$A(q^2)$ is independent of the leptonic couplings.
Therefore we can evaluate the above asymmetry without
any theoretical uncertainty in the heavy quark limit.

\section{Implication of the heavy quark limit}
Here, we discuss the implication of the heavy quark limit
on the asymmetry $A(q^2)$. In general, we cannot evaluate
$A(q^2)$ because it depends on the unknown form factors
defined in eqs.~(\ref{VEC}) and (\ref{AX}).
This difficulty, however, greatly reduces
in the heavy quark limit.

In the limit that $m_b,m_c\rightarrow \infty$, the six form
factors in eqs.~(\ref{VEC}) and (\ref{AX}) can be written
by only one unknown function of $q^2$ \cite{IWMRR}:
\begin{equation}
F_1=G_1\equiv\zeta(q^2),\;F_2=F_3=G_2=G_3=0\,.
\label{HQL}
\end{equation}
As seen in the above equation (\ref{HQL}), the spin of
heavy $\Lambda_Q$ hyperon is carried by the heavy quark. This is
also true for $Z\rightarrow b\bar b$ followed by
$b\rightarrow \Lambda_b$. Therefore, we can use
eq.~(\ref{P}) as the polarization of the $\Lambda_b$
which comes from $Z$ decay in the heavy $b$ quark limit.

Using eq.~(\ref{HQL}), (\ref{J}) and (\ref{K}),
we get
\begin{equation}
J(q^2)=\frac{\vert f_V\vert ^2+\vert f_A\vert ^2}
            {192\pi^3\beta E_b}\zeta (q^2)^2
      \left[\vert g_V\vert ^2(w_+w_-+3 q^2w_-)
            +\vert g_A\vert ^2(w_+w_-+3 q^2w_+)\right]\;,
\end{equation}
\begin{equation}
K(q^2)=\frac{\vert f_V\vert ^2+\vert f_A\vert ^2}
            {96\pi^3\beta^2 E_b}\zeta (q^2)^2
       (g_Vg_A^*+g_V^*g_A)m_{\Lambda_b}^2
        (m_{\Lambda_b}^2-m_{\Lambda_c}^2-2 q^2)\;,
\end{equation}

Then, according to eq.~(\ref{A}),
we get an expression for $A(q^2)$ in the heavy quark limit:
\begin{equation}
A(q^2)=-{\cal P}
       \frac{(\vert g_L \vert^2-\vert g_R \vert^2)
             (m_{\Lambda_b}^2-m_{\Lambda_c}^2-2q^2)\sqrt{w_+w_-}}
 {(\vert g_L \vert^2\!+\!\vert g_R\vert^2)
   \left\{2w_+w_-+3q^2(w_++w_-)\right\}
   -3(g_Lg_R^*\!+\!g_L^*g_R)q^2(w_+-w_-)}\;,
\label{AR}
\end{equation}
where $g_L=g_V-g_A\,,\;g_R=g_V+g_A$. As is mentioned above,
this expression has no theoretical uncertainty if $g_R/g_L$
is given. The numerical result of this
expression is shown in fig.~\ref{FIGHQL}. Throughout this paper, we
use $m_{\Lambda_c}=2.285$GeV \cite{PDG} and $m_{\Lambda_b}=5.640$GeV
\cite{UA1} in numerical calculations.

In fig.~\ref{FIGHQL}, $A(q^2)$ is plotted for pure left-handed case
($g_R/g_L=0$), 30\% right-handed contamination cases ($g_R/g_L=\pm
0.3$) and pure right-handed case ($g_L/g_R=0$). In the cases of 30\%
right-handed contamination, it is assumed that the left-handed
coupling and the right-handed one are relatively real and the two
relative sign possibilities are taken into account. Fig.~\ref{FIGHQL}
shows that we can easily discriminate between the pure left-handed
case and the pure right-handed case by measuring $A(q^2)$ for smaller
$q^2$.  An experiment with enough accuracy will give an upper bound
for the strength of right-handed current, or will find a right-handed
current.

{}From eq.~(\ref{AR}), we can immediately find that $A(q^2)$ takes the
following values at the kinematical points of zero momentum transfer
and zero recoil:
\begin{equation}
A(0)=-\frac{1}{2}{\cal P}\frac{\vert g_L \vert^2-\vert g_R \vert^2}
{\vert g_L \vert^2+\vert g_R \vert^2}\,,
\label{A0}
\end{equation}
\begin{equation}
A((m_{\Lambda_b}-m_{\Lambda_c})^2)=0\,.
\end{equation}

At this stage, we conclude that we can clearly distinguish the
left-handed case from the right-handed case by measuring
the asymmetry $A(q^2)$ near the kinematical point of
zero momentum transfer, at least in the heavy quark limit.
It should be noticed that the phase
space volume is finite at the zero momentum transfer, while it
vanishes at the point of zero recoil.

\section{$1/{\lowercase{m_c}}$ correction}
In this section, we discuss the $1/m_c$ correction to the result of
the previous section. The charm quark mass is not so heavy compared
with a typical QCD scale that the $1/m_c$ correction is
expected to be the dominant one, and one should
take it into account.

Including the leading $1/m_c$ correction, eq.~(\ref{HQL}) is modified
as follows \cite{GGW}:
\begin{equation}
F_1=(1+\Delta)\zeta(q^2)\,,\;G_1=\zeta(q^2)\,,
\;F_2=G_2=-\Delta\,\zeta(q^2)\,,\;
F_3=G_3=0\,,
\label{MC}
\end{equation}
where
\begin{equation}
\Delta=\frac{\bar\Lambda}{m_c}\left (\frac{1}{1+v\cdot v'}\right )\,,
\end{equation}
and $\bar\Lambda=m_{\Lambda_b}-m_b=m_{\Lambda_c}-m_c$. $\bar\Lambda$
is an unknown parameter and is estimated as
$\bar\Lambda\sim 0.7$GeV.
Note that the form factors are still proportional to one unknown
function $\zeta(q^2)$.  This means that the asymmetry $A(q^2)$ does
not involve any unknown functions even with the $1/m_c$ correction
although it depends on the unknown parameter $\bar\Lambda$.

Using eq.~(\ref{MC}), one can write down an analytic formula for
$A(q^2)$ in terms of $\bar\Lambda$. It, however, is too lengthy to be
presented here. We show only the numerical result
in fig.~\ref{FIGMC}.

In fig.~\ref{FIGMC}, $A(q^2)$ for the pure left-handed case and the
pure right-handed case are plotted with and without $1/m_c$
correction. ($\bar\Lambda=0.7$GeV and $\bar\Lambda=0$ respectively.)
This figure shows that the $1/m_c$ correction decreases quickly with
approaching to zero momentum transfer and vanishes at the point of
zero momentum transfer.

Actually, it is easy to demonstrate the latter analytically.
At the point $q^2=0$,
we get the same result as eq.~(\ref{A0}) even in the presence of
$1/m_c$ correction. Using eqs.~(\ref{J}), (\ref{K}) and
(\ref{MC}), one gets
\begin{equation}
J(0)=\frac{\vert f_V \vert^2+\vert f_A \vert^2}{384\pi^3\beta E_b}
     \zeta (0)^2 (\vert g_L \vert^2+\vert g_R \vert^2)
     (m_{\Lambda_b}^2-m_{\Lambda_c}^2)^2
     \left\{1+\Delta_0\left(1-m_{\Lambda_c}/m_{\Lambda_b}\right)
     \right\}\;,
\label{J0}
\end{equation}
\begin{equation}
K(0)=-\frac{\vert f_V \vert^2+\vert f_A \vert^2}{192\pi^3\beta^2 E_b}
     \zeta (0)^2 (\vert g_L \vert^2-\vert g_R \vert^2)
     m_{\Lambda_b}^2(m_{\Lambda_b}^2-m_{\Lambda_c}^2)
     \left\{1+\Delta_0\left(1-m_{\Lambda_c}/m_{\Lambda_b}\right)
     \right\}\;,
\label{K0}
\end{equation}
where $\Delta_0=\Delta\vert_{q^2=0}$ and
we neglected the terms of higher order in $\Delta_0$.
Using eqs.~(\ref{A}), (\ref{J0}) and (\ref{K0}), one finds the
same result of eq.~(\ref{A0}) \cite{KK}.

{}From the above argument, we conclude that the $1/m_c$ correction can
be negligible if one does not go near the zero recoil point
($q^2=(m_{\Lambda_b}-m_{\Lambda_c})^2$). Especially, it vanishes at
the point of zero momentum transfer ($q^2=0$).

\section{Conclusion}
Before summarizing our result, two remarks are in order:
\begin{enumerate}
\item {\it Effect of fragmentation}: One needs to know the
       momentum of the initial $\Lambda_b$ to measure the asymmetry
$A(q^2)$.  If the initial $b$ quark hadronizes into a $\Lambda_b$
without other hadrons, the absolute value of the $\Lambda_b$ momentum
is simply given by $\sqrt{(m_Z/2)^2-m_{\Lambda_b}^2}\;$.  In general,
however, $b$ quarks in $Z$ decay hadronize with several hadrons, {\em
i.~$\!\!$e.~} they form jets.  Then the magnitude of the $\Lambda_b$
momentum is no longer a constant. It varies according to some
fragmentation function $D(z)$ where $z$ is the energy or momentum
fraction carried by $\Lambda_b$ \cite{FOOT2}.  This affects a
measurement of the double differential rate of eq.~(\ref{DDR}).\\
\hspace*{2em}To discuss the fragmentation effect, we consider the
model of Peterson {\em et~al.~} \cite{P}.  In this model, the
fragmentation function is given by
\begin{eqnarray}
D(z)&=&Nz^{-1}
\left(1-\frac{1}{z}-\frac{\epsilon}{1-z}\right)^{-2}\,, \label{PF}
\end{eqnarray}
where $N$ is a normalization factor and $\epsilon$ is a
parameter.  We can write $\epsilon$ as $\epsilon\simeq m_q^2/m_Q^2$,
where $m_q$ is the mass of light quark and $m_Q$ is that of the heavy
quark. Eq.~(\ref{PF}) describes the experimental data of $\Lambda_c$
production well \cite{ACC}, and the scaling of $\epsilon$ as
$m_Q^{-2}$ has been observed by comparing charm and bottom
fragmentation \cite{Chrin}. Therefore, the use of
this model for a qualitative discussion seems to
be legitimate.\\ \hspace*{2em}The maximum value of $D(z)$ in
eq.~(\ref{PF}) is given at \begin{eqnarray}
z_{max}&=&1+\frac{\epsilon}{2}-
\frac{1}{2}\sqrt{\epsilon(\epsilon+4)}\\
&=&1-\sqrt{\epsilon}+\cdots\,.  \label{ZMAX} \end{eqnarray} Therefore
the fragmentation effect seems to be $1/m_b$ effect in $\Lambda_b$
production. Since we ignored $1/m_b$ effect in the arguments of the
previous sections, the fragmentation effect can be neglected,
at least in the formal point of view. \\
\hspace*{2em}While the average of $z$
which is distributed according to the fragmentation function in
eq.~(\ref{PF}) cannot be expanded in $\sqrt{\epsilon}$:
\begin{eqnarray} \left\langle z\right\rangle&=&1+
\frac{2}{\pi}\sqrt{\epsilon}\log\epsilon+\cdots\,.  \end{eqnarray}
This equation suggests that the effect of fragmentation is not so
small. To be more precise, Monte Carlo study seems to be needed to
extract the asymmetry in eq.~(\ref{AD}) from experimental data. This
is beyond the scope of the present paper.\\
\item {\it $\mit\Lambda_b$ from polarized $e^+e^-$ collision
       near threshold}: To get rid of the fragmentation effect, one
may produce $\Lambda_b$ in $e^+e^-$ collision near
$\Lambda_b\bar\Lambda_b$ threshold.  In this case, however, polarized
beams are needed to get the necessary polarization of $\Lambda_b$.
In the collision of right-handed electron and left-handed positron,
the polarization of $b$ quark is given by
\begin{eqnarray}
{\cal P}&=&\frac{2 s \cos \theta}
         {s(1+\cos^2 \theta)+4 m_b^2(1-\cos^2\theta)}\,,
\label{PPC}
\end{eqnarray}
where $s$ is the center-of-mass
energy squared, and $\theta$ denotes the angle between the direction
of the incoming electron momentum and the that of the outgoing $b$
quark. One can use eq.~(\ref{PPC}) for $\Lambda_b$ polarization if
$1/m_b$ correction is neglected.\\ \hspace*{2em}If one approaches
close to the threshold, the produced $\Lambda_b$ is almost stopped.
In  the rest frame of $\Lambda_b$,
the asymmetry defined in eq.~(\ref{AD})
has no meanings.  However, we can observe the same physical effect,
{\em i.~$\!\!$e.~} the spin-momentum correlation, by measuring the
angular distribution of $\Lambda_c$.
\end{enumerate}

To summarize, we discussed $\Lambda_b\rightarrow\Lambda_c\ell\bar\nu$
decay with an initial polarization using the heavy quark effective
theory.  We defined an asymmetry which was sensitive to the chirality
of the $b\rightarrow c$ charged current and independent of that of
leptonic charged currents. In the heavy quark limit, this asymmetry
has no theoretical uncertainty.  The $1/m_c$ correction to it is
negligibly small if one does not go near the kinematical point of
zero recoil. Moreover, the $1/m_c$ correction vanishes at the point
of zero momentum transfer.

According to the above result, we conclude that the investigation of
the semileptonic $\Lambda_b$ decay with an initial polarization will
provide a good test of the left-handedness of the $b\rightarrow c$
current and lead to a limit or an evidence on existence of a
right-handed $b\rightarrow c$ current with satisfactory theoretical
accuracy.

\acknowledgments

The author would like to thank Prof.~Hikasa, Prof.~K\"orner and
Prof.~Wakaizumi for useful discussions.

\unletteredappendix{%
  Explicit forms of $J(\lowercase{q^2})$ and $K(\lowercase{q^2})$}

Here, we give the explicit forms of $J(q^2)$ and $K(q^2)$
which appear in eq.~(\ref{DDR}).
\begin{eqnarray}
J(q^2)&=&\frac{\vert f_V\vert ^2+\vert f_A\vert ^2}
            {192\pi^3\beta E_b}\nonumber\\
      & & \times\biggl[\vert g_V\vert ^2
             \left[w_+ w_-\left\{F_1^2+
                           2(m_{\Lambda_b}+m_{\Lambda_c})F_1F_+
                           +w_+F_+^2\right\}+
                   3q^2w_-F_1^2\right]\nonumber\\
      & &  {}+\vert g_A\vert ^2
             \left[w_+ w_-\left\{G_1^2-
                           2(m_{\Lambda_b}-m_{\Lambda_c})G_1G_+
                           +w_-G_+^2\right\}+
                   3q^2w_+G_1^2\right]\biggr]\label{J}\;,
\end{eqnarray}

\begin{eqnarray}
K(q^2)&=&\frac{\vert f_V\vert ^2+\vert f_A\vert ^2}
            {96\pi^3\beta^2 E_b}(g_Vg_A^*+g_V^*g_A)
          m_{\Lambda_b}^2\nonumber\\
    & &\times\Bigl[(m_{\Lambda_b}^2-m_{\Lambda_c}^2-2 q^2)F_1G_1
                  -(m_{\Lambda_b}+m_{\Lambda_c})w_-F_1G_+
                  \nonumber\\
           & &{}+(m_{\Lambda_b}-m_{\Lambda_c})w_+F_+G_1
                  -w_+w_-F_+G_+\Bigr]\label{K}\;,
\end{eqnarray}
where $f_V$ and $f_A$ denote the leptonic vector and
axial-vector couplings,
$F_+=(F_2/m_{\Lambda_b}+F_3/m_{\Lambda_c})/2$ and
$G_+=(G_2/m_{\Lambda_b}+G_3/m_{\Lambda_c})/2$.
Note that only $F_1$, $G_1$ and the combinations $F_+$
and $G_+$ survive in the above expressions because of
vanishing lepton masses.

\figure{$A(q^2)$ in the heavy quark limit for pure left-handed
        case (solid line), 30\% right-handed contamination cases
        (dashed and dash-dotted lines), and pure right-handed
        case (dotted line).\label{FIGHQL}}
\figure{$A(q^2)$ for pure left-handed case with (dashed line) and
         without (solid line) $1/m_c$ correction, and that for pure
         right-handed case with (dash-dotted line) and without
         (dotted line) $1/m_c$ correction.\label{FIGMC}}


\begin{references}
\bibitem{FOOT1}
 Some trivial factors like $G_F$ may be
 absorbed into these couplings.
\end{references}
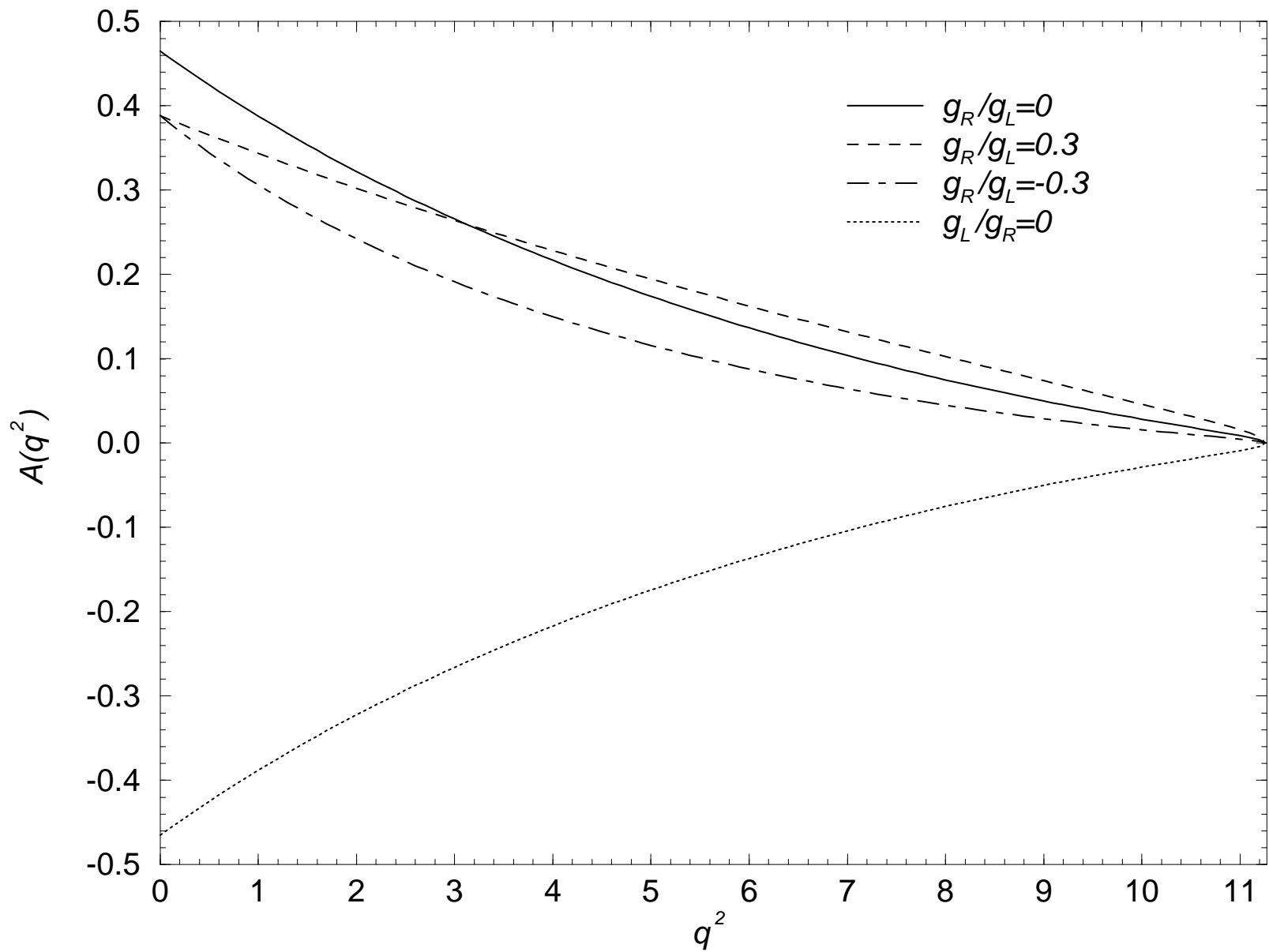
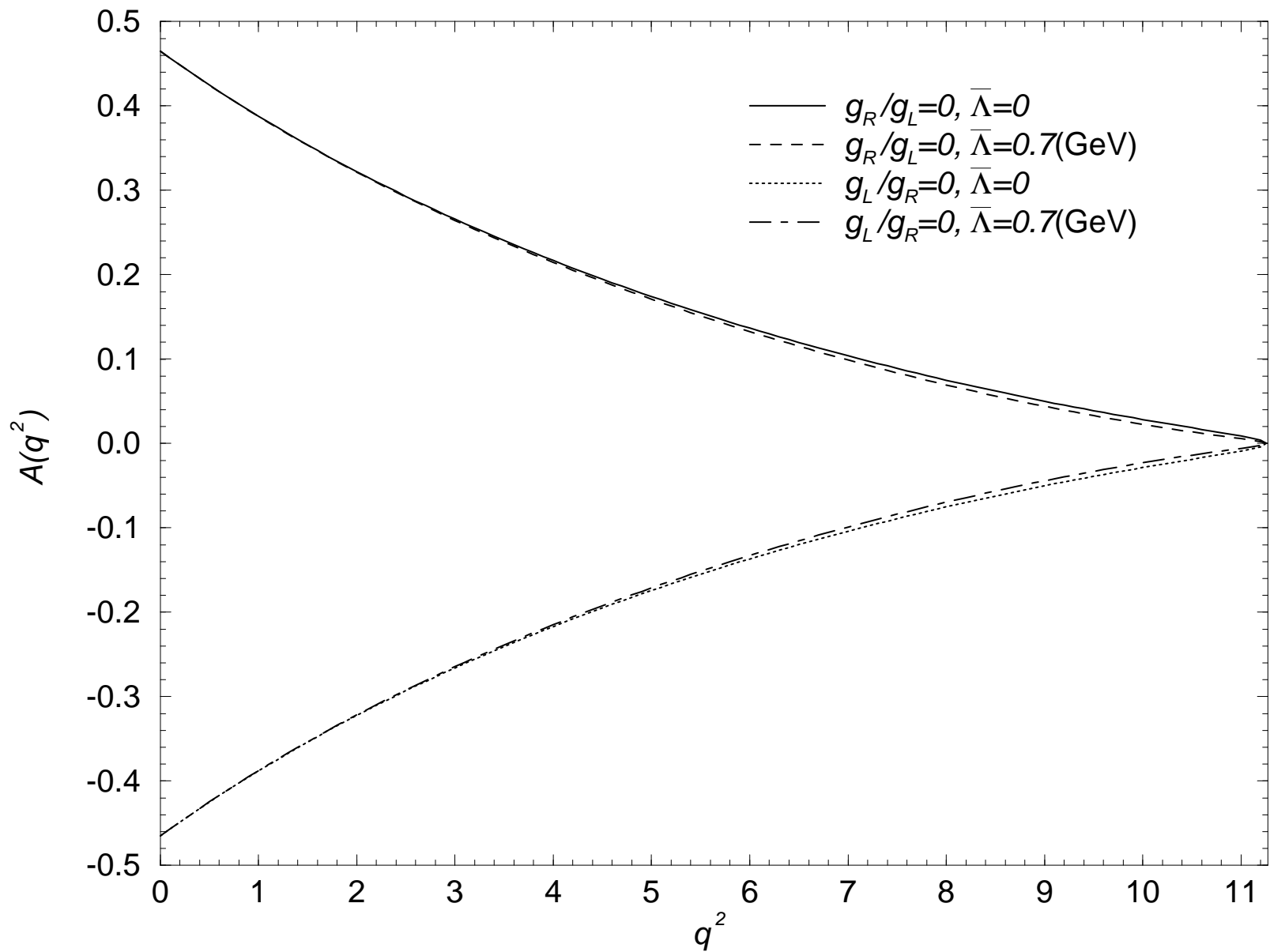
\end{document}